\documentclass[twocolumn,secnumarabic,amssymb, nobibnotes, aps,notitlepage]{revtex4-1}
\usepackage{graphicx}
\usepackage{tabularx} 
\usepackage{hyperref}
\usepackage{amsmath}
\usepackage{xcolor}
\usepackage{multirow}
\newcommand{\fourpi}{$\pi^+\pi^-\pi^+\pi^-$}
\begin{document} 

\title{Photoproduction and detection of $\rho'\rightarrow\pi^+\pi^-\pi^+\pi^-$ decays in ultra-peripheral collisions and at an electron-ion collider}

\author{Neha Devi} \altaffiliation{Current address:  Jaeger Corporation, P.O. Box 540364
Omaha, NE 68154}

\affiliation{Creighton University, 2500 California Plaza, Omaha, NE 68178 USA}

\author{Minjung Kim}
\affiliation{Nuclear Science Division, Lawrence Berkeley National Laboratory, 1 Cyclotron Road, Berkeley, CA 94720 USA}
\affiliation{Department of Physics University of California, 366 Physics North MC 7300, Berkeley, CA 94720 USA}
\affiliation{Center for Frontiers in Nuclear Science, Stony Brook University, Stony Brook, NY 11794 USA}
\author{Spencer R. Klein}
\email[]{srklein@lbl.gov} 
\affiliation{Nuclear Science Division, Lawrence Berkeley National Laboratory, 1 Cyclotron Road, Berkeley, CA 94720 USA}
\author{Janet Seger}
\affiliation{Creighton University, 2500 California Plaza, Omaha, NE 68178 USA}

\date{\today}
\begin{abstract}

Vector meson photoproduction is an important probe of  nuclear structure.  Light vector mesons are most sensitive to low$-x$ structure, as long as they are not too light for perturbative QCD calculations.   The $\rho'$ is of interest as an intermediate mass state (between the $\rho$ and $J/\psi$) that is easier to detect than the $\phi$. 

Using HERA data on proton targets, we make projections for lead/gold targets in UPCs at the Large Hadron Collider and RHIC, and for $ep$ and $eA$ collisions at a future Electron-Ion Collider (EIC).   We compare the UPC projections with ALICE data, and constrain the branching ratio divided by the square of the photon-$\rho'$ coupling.  The data prefer large couplings and small branching ratio, probably less than 25\%.  The photon-meson coupling predicted by generalized vector meson dominance does not fit the data.  The HERA $ep$ and ALICE UPC $e$Pb data exhibit very similar $4\pi$ mass spectra, indicating that, if the system is composed of two resonances, the products of their photon couplings with their four-pion branching ratios are similar.

The predicted rates are high for both UPCs and the EIC.  The $\rho'\rightarrow\pi^+\pi^-\pi^+\pi^-$ decay can be observed at the EIC with high efficiency.  In $ep$ collisions at the highest energy, the forward B0 detector is needed to observe this channel down to the lowest achievable Bjorken$-x$ values.

\end{abstract}
\maketitle

\section{Introduction}

Vector meson photoproduction has been studied extensively at fixed-target accelerators \cite{Bauer:1977iq}, the HERA $ep$ collider \cite{Ivanov:2004ax}, and with ultra-peripheral collisions (UPCs) at heavy-ion colliders \cite{Klein:2020fmr}.   It will also be an important probe  of nucleons and nuclei at a future electron-ion collider (EIC)  \cite{Accardi:2012qut,AbdulKhalek:2021gbh,ATHENA:2022hxb}. The Good-Walker paradigm relates coherent vector meson production to the  average nuclear configuration, while incoherent vector meson photoproduction is related to fluctuations in the nuclear configuration, including gluonic hotspots
\cite{Miettinen:1978jb,Mantysaari:2016ykx,Klein:2019qfb}.  

The $Q^2$ dependence of exclusive production is an important signature of saturation  \cite{Mantysaari:2017slo}.  Definitive conclusions about saturation will require studies of different mesons, with different wave functions and masses.  

The $\rho$ is straightforward to reconstruct \cite{STAR:2002caw}, but from the theory perspective is rather light, limiting the applicability of perturbative QCD (pQCD) based calculations.  The $J/\psi$ is heavy enough that saturation phenomena are greatly reduced \cite{Mantysaari:2017slo}. The $\phi$ is attractive because it has an intermediate mass (between the $\rho$ and the $J/\psi$).

Early plans for exclusive vector meson production at a U. S. EIC focused on the $\phi$ and $J/\psi$ \cite{Accardi:2012qut}.  However, $\phi$ production at low $Q^2$ is hard to reconstruct \cite{AbdulKhalek:2021gbh,Arrington:2021yeb} because the main channel, $\phi\rightarrow K^+K^-$,   suffers from a low $Q$ value, with the daughter kaons having a momentum of only 127 MeV/c in the $\phi$ rest frame.  Other final states have either low branching ratios or include a long-lived $K^0_L$.  For UPCs, the situation is similar, with coherent $\phi$ photoproduction difficult to observe \cite{ALICE:2023kgv,CMS:2025lsm}.  

The $\rho'$ states are attractive alternatives to the $\phi$, as they also have intermediate masses, between the $\rho$ and $J/\psi$.   However, these states have a more complex wave function and a more complex phenomenology.  There are likely two overlapping resonances, the $\rho'(1450)$ and the $\rho'(1700)$ \cite{ParticleDataGroup:2022pth}.  These resonances can decay to many different final states, but both have a significant branching ratio to \fourpi.  Since the \fourpi final state is easy to reconstruct, we will focus on it.  Most photoproduction analyses have fit the \fourpi mass spectrum to a single resonance, so we will perforce do the same here. 

The $4\pi$ final state has been studied at fixed-target accelerators, using $ep$ collisions at HERA and in ultra-peripheral collisions (UPCs) of gold ions at RHIC and lead ions at the LHC \cite{ParticleDataGroup:2024cfk}.  Using HERA and fixed-target data on proton targets as input, we will use a Glauber calculation to predict the RHIC and LHC cross sections for ion targets.  

We also make predictions about the the $\rho'$ cross sections in $ep$ and $eA$ collisions at the EIC, and estimate the reconstruction efficiency using a simple model of the proposed ePIC detector. As with UPCs, the $eA$ cross sections depend on the $4\pi$ branching ratio. 

\section{Modeling of $\rho'$ photoproduction}

The $\rho'$ states are radial excitations of the $\rho$, with the same $J^{PC}=1^{--}$ quantum numbers.   The first studies of $4\pi$ photoproduction were done in fixed-target experiments at photon energies from 3 GeV to 70 GeV\cite{Bingham:1972az,Schacht:1974fq,Alexander:1975xe,LAMP2Group:1979ibr,Bonn-CERN-EcolePoly-Glasgow-Lancaster-Manchester-Orsay-Paris-Rutherford-Sheffield:1981ich,OmegaPhoton:1984rzb,Atiya:1979ip}.   Four pion photoproduction was first studied at collider energies by the Solenoidal Tracker at RHIC (STAR) collaboration \cite{STAR:2009giy}, and later by the ALICE detector at the LHC \cite{ALICE:2024kjy}.   
The H1 collaboration has also studied 4-pion photoproduction, and observed a resonance with similar parameters to those measured by STAR.  Both STAR and H1 found that the data was  well fit by a single resonance \cite{Schmitt,H1prelim}. We characterize the resonance using the STAR results: mass $M_V=1570$\ MeV$/c^2$, and width, $\Gamma=570\pm 60\ $MeV$/c^2$, with a momentum-dependent width.  We use the H1 cross section data as input to make predictions for ion targets in UPCs and at the EIC.  

\subsection{Proton targets}

We fit H1 and fixed-target data on $\sigma(\gamma p\rightarrow \rho'p\rightarrow \pi^+\pi^-\pi^+\pi^- p)$ to a two-component Reggeon $+$ Pomeron model \cite{Klein:1999qj, Klein:2016yzr}:  
\begin{align}
\sigma(\gamma p\rightarrow \rho'p\rightarrow \pi^+\pi^-\pi^+\pi^- p) = (XW^\epsilon+YW^{-\eta})
\nonumber
\\
\cdot{\rm Br}(\rho'\rightarrow \pi^+\pi^-\pi^+\pi^-).
\label{eq:pomeronreggeon}
\end{align}
Since the branching ratios are unknown, the data are for the cross sections times branching ratio.  

\begin{figure}[tb]
    \centering
    \includegraphics[width=0.48\textwidth]{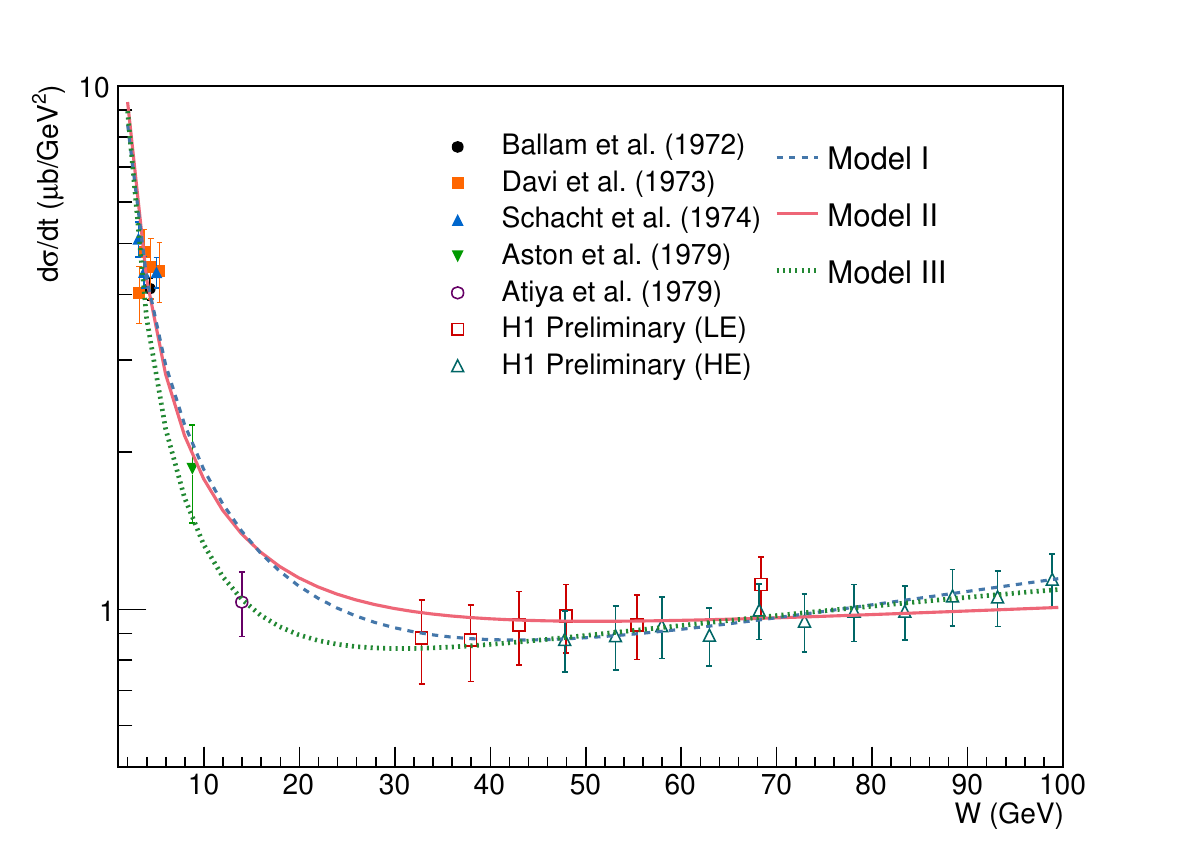}
    \caption{H1 \cite{Schmitt,H1prelim} and fixed-target data \cite{Bingham:1972az, Davier1973, Schacht:1974fq, Atiya:1979ip, Aston1981} on $\gamma p\rightarrow \rho'p\rightarrow\pi^+\pi^-\pi^+\pi^-$ along with our fit to the data. Model I is a fit to Eq. \ref{eq:pomeronreggeon} with all four parameters treated as free parameters.    Model II is the reference fit used for the rest of the calculations in this paper. This fit has an additional term in the $\chi^2$, pulling $\epsilon$ toward the value from H1 \cite{Schmitt,H1prelim},as discussed in the text.  Model III uses the set of parameters used in Ref.  \cite{Klusek-Gawenda:2020gwa}. Fit results are given in Tab. \ref{tab:fitparam}.}
   \label{fig:H1fit}
\end{figure}

The first term represents Pomeron exchange (with strength $X$ and power-law exponent $\epsilon$), while the second is for Reggeon exchange (with strength $Y$ and power-law exponent $-\eta$).  Br is the branching ratio for the specified decay.   Figure \ref{fig:H1fit} shows the data, with both statistical and systematic errors.  The H1 systematic uncertainties  include point-to-point correlations.   Also shown are three fits to the data, with the parameters given in Tab. \ref{tab:fitparam}.  These parameters assume a  branching ratio of 100\% for $\rho'\rightarrow \pi^+\pi^-\pi^+\pi^-$.   For smaller branching ratios, $X$ and $Y$ should be scaled up by dividing by the branching ratio. 

In Model I, all four parameters in Eq. \ref{eq:pomeronreggeon} were allowed to float.  This led to a fairly large value of $\epsilon$ of 0.87, considerably above expectations for Pomeron exchange.  In this fit, the Pomeron contribution was generally small, with the cross section dominated by the Reggeon component, even at H1 energies.
 
The second fit, Model II, added a nuisance term, based on the value for $\epsilon$ found by H1.  A term $(\epsilon-\epsilon_{H1})^2/\sigma^2_{H1}$ was added to the $\chi^2$ in the fit, where $\epsilon_{H1}=0.23$ and $\sigma_{H1}=0.06$ are taken from the fit in 
Fig. 4 of Ref. \cite{H1prelim}.  The rationale for this was that $\sigma_{H1}$ was dominated by systematic errors;  the H1 fit correctly accounted for the correlated systematic errors in their analysis, so was more precise than if we just used their data points and errors.  This fit pulled $\epsilon$ to 0.28, slightly larger than that found by H1, and many other studies of light mesons \cite{Szuba:2009zz}.   The downside is that this approach uses the H1 data twice - in the H1 $\epsilon$ and in our fit.  This decreases the importance of the fixed-target data.  Since that data is far away in energy from the region of interest, this should not cause problems.  
 
The third fit, Model III, is from Ref. \cite{Klusek-Gawenda:2020gwa}.   It differs from our fit in that it includes a slightly larger selection of fixed-target data, especially at very low energies.   It found an $\epsilon$ intermediate between Model I and Model II.

Models I and III both have large values of $\epsilon$, and so, at large energies, the cross section will grow overly rapidly, unlike other light vector mesons.  They also have a rather large fractional Reggeon contribution ($X/Y$ is small), in tension with the H1 fit and expectations based on other vector mesons.  For these reasons, the remainder of the paper uses the Model II fit. 

The differences in predictions between the three fits for energies above 20 GeV are generally within 15\%. The $\chi^2/ndf$ for Model 2 is 1.736, which is marginally higher than that of Model 1 (1.440) and significantly lower than that of Model 3 (3.049).

\begin{table}
    \centering
     \begin{tabularx}{\linewidth}{c|>{\centering\arraybackslash}X>{\centering\arraybackslash}X>{\centering\arraybackslash}X>{\centering\arraybackslash}X}
     \
         & $X (\mu {\rm b})$ & $\varepsilon$ & $Y (\mu {\rm b})$ & $\eta$\\
         \hline\hline
        Model I & 0.02 & 0.87 & 16.61 & 0.99 \\
        \hline
         Model II& 0.26 & 0.28  & 20.78 & 1.21  \\
        \hline
        Model III  & 0.16 & 0.41 & 23.0 & 1.4\\
    \end{tabularx}
    \caption{Fit parameters, assuming a 100\% branching ratio for $\rho'\rightarrow\pi^+\pi^-\pi^+\pi^-$.  These fits results are for 100\% branching ratio.  For other branching ratios, $X$ and $Y$ should be divided by the branching ratio.}
    \label{tab:fitparam}
\end{table}

The $\rho'$ branching ratio in Eq. \ref{eq:pomeronreggeon} 
converts the $\rho'$ cross section into the $\pi^+\pi^-\pi^+\pi^-$ cross section.  The branching fraction is important for ion targets, since a Glauber calculation, discussed below, does a non-linear mapping of $\sigma(\gamma p\rightarrow \rho' p)$ into $\sigma(\gamma A\rightarrow \rho' A)$, the cross section on an ion target.  Because of the non-linearity, it is necessary to remove the branching ratio before this mapping.

The branching ratio to $\pi^+ \pi^- \pi^+ \pi^-$ is poorly known.  Ref.~\cite{Frankfurt:2002sv}  suggested $Br(\rho'\rightarrow \pi^+ \pi^- \pi^+ \pi^-\approx 30\%$, based on the inferred cross-section for $\rho'$ decaying into $4\pi$ in gold-gold collisions at RHIC, about one-third of that for the $\rho$.  Another study, estimated a branching ratio of $\approx 40\%$
(summing nonresonant $\pi^+\pi^-\pi^+\pi^-$ and $\rho^0\pi^+\pi^+$) \cite{H1:2020lzc}.
We consider branching ratios between 10\% and 100\%.  100\% is the obvious maximum, while a branching ratio below 10\% would require that the $\rho'$ cross section is larger than that for the $\rho^0$, which seems unlikely.  

\subsection{Production on ion targets}

The cross section for photoproduction on nuclear targets is found using a quantum Glauber calculation \footnote{This differs from Ref.
\cite{Klusek-Gawenda:2020gwa} which used a classical Glauber calculation}.  The cross section for forward production, $d\sigma/dt|_{t=0}$ is equal to $b_V\sigma(k)$, where $b_V$ is the slope of $d\sigma/dt$ at small $t$.  This forward cross section has two parts: the probability for a photon to fluctuate into the $\rho'$, and the $\rho'$-nucleon elastic scattering cross section:
\begin{equation}
\frac{d\sigma(\gamma p\rightarrow Vp)}{dt}\bigg|_{t=0} =
\frac{4\pi\alpha}{f_V^2}
\ \ 
\frac{d\sigma(Vp\rightarrow Vp)}{dt} \bigg|_{t=0} .
\end{equation}
We take $b_v=9.4 \pm 0.3 \text{ stat } \pm 1.0 \text{ sys}$ GeV$^{-2}$, taken from the coherent  (softer) exponent in Fig. 6 of Ref. \cite{H1prelim} to convert from the total cross section to $d\sigma/dt$.

The vector-meson photon coupling $f_V$ is usually determined from the meson coupling to $e^+e^-$: 
\begin{equation}
    \frac{f_V^2}{4\pi}=\frac{M_V\alpha^2}{3\Gamma_{V\rightarrow ee}},
\end{equation}
where $M_V$ is the vector meson mass, $\Gamma_{V\rightarrow ee}$ is the partial width for that meson to decay to $e^+e^-$ and $\alpha$ is the fine structure constant.  The coupling $f_V$ is unmeasured for the $\rho'$. 

As will be shown, the ratio of proton-target to ion-target cross sections are sensitive to 
$\Gamma_{V\rightarrow ee}$.  There are two prior estimates. Ref. \cite{Klusek-Gawenda:2020gwa} estimated $\Gamma_{V\rightarrow ee} = 0.425\pm 0.075$ keV and $f_V^2/4\pi \approx 65.6$ for the $\rho(1570)$. Alternately, Generalized Vector Meson Dominance (GVMD) predicts that \cite{Frankfurt:1997zk}
\begin{equation}
    \frac{f_V^2}{f_{\rho^0}^2} = \frac{M^2_{\rho(1570)}}{M^2_{\rho(770)}}
    \label{eq:GVMD}.
\end{equation}
This leads to $f_V^2/4\pi \approx 8.4$ and $\Gamma_{V\rightarrow ee}=1.76$ keV, or about four times greater than Ref. \cite{Klusek-Gawenda:2020gwa}.  This GVMD treatment lacks off-diagonal elements that couple different mesons, which may be important for $\rho'$ \cite{Bronstein:1977qe}.  The GVMD estimate of $f_V^2$ also leads to a surprisingly small cross section, as will be discussed below.   The use of two resonances, each with the GVMD coupling, would not reduce these discrepancies.    For these reasons, we use the coupling from Ref. \cite{Klusek-Gawenda:2020gwa} as our baseline, while considering other possibilities.

The optical theorem is then used to find the total $\rho' p$ cross section. 
\begin{equation}
\sigma_{Tot}^2 (Vp) = 16\pi 
\frac{d\sigma(Vp\rightarrow Vp)}{dt} \bigg|_{t=0}.
\end{equation}
At $W=10$ GeV, $\sigma_{Tot}=18$ mb, assuming a 100\% branching ratio, and 57 mb for a 10\% branching ratio.   In comparison, the cross sections for the $\rho$ and $\omega$ are 24 and 26 mb respectively \cite{Klein:1999qj}.  It seems unlikely that the $\rho'$ cross section is twice that of the $\rho$.  This disfavors scenarios with low branching ratios. 

The $VA$ cross section is then found using a quantum Glauber calculation \cite{Frankfurt:2002sv}:
\begin{equation}
    \sigma_{Tot}(VA) = \int d^2\vec{r}\ 2\big(1-e^{-\sigma_{Tot}(Vp)T_A(\vec{r})/2}\big),
\end{equation}
where $T_A(\vec{r})$ is the nuclear thickness function.  

The optical theorem can be used again, to convert $\sigma_{Tot}(VA)$ into $d\sigma(\gamma A\rightarrow \rho' A)/dt|_{t=0}$. Then, we use the nuclear form factor $F(q)$ from Ref. \cite{Klein:1999gv} to get the total $\rho'$ cross-section.  
\begin{align}
\sigma(\gamma A\rightarrow \rho' A\rightarrow\pi^+\pi^-\pi^+\pi^- A) = 
    \frac{\sigma\!(\!\gamma\! A\!\rightarrow\! \rho'\! A\!)}{dt} \bigg|_{t=0}\! 
    \nonumber
\\
\int_{t_{\rm min}}^\infty \! dt |F(t)|^2 \cdot{\rm Br}(\rho'\rightarrow \pi^+\pi^-\pi^+\pi^-).
\label{eq:rhoingammaA}
\end{align}

This is then combined with the photon flux in UPCs from Ref. \cite{Klein:1999gv} or for $ep/eA$ collisions in Ref. \cite{Lomnitz:2018juf} to find the cross sections for UPC or $ep/eA$ collisions. 

For PbPb collisions at LHC energies, using the two different estimates for $f_V^2$ results in cross sections that differ by about a factor of four. Varying the branching ratio from 10\% to 100\% changes the cross section by approximately a factor of five.  In comparison, the 15\% uncertainty from the choice of $\gamma p\rightarrow 4\pi p$ is negligible.   As we will see, ALICE data significantly narrows the possible combinations.

\section{$\rho'$ production in ultra-peripheral collisions}

The cross section for photoproduction in UPCs is given by combining the $\gamma p$ (Eq. \ref{eq:pomeronreggeon}) or $\gamma A$ cross sections (Eq. \ref{eq:rhoingammaA}) with the photon flux for protons \cite{Klein:2003vd} or ions \cite{Klein:1999gv}.   This code is implemented in the STARlight Monte Carlo code \cite{Klein:2016yzr}.

Figure \ref{fig:BRmodels} shows the calculated $d\sigma/dy$ for the $\rho'$ for possible $4\pi$ branching ratio between 10\% and 100\% and for the two different possible $f_v$s.  The calculations are compared with ALICE data on 5.02 TeV lead-lead collisions \cite{ALICE:2024kjy}. Unfortunately, the STAR data on the $\pi^+\pi^-\pi^+\pi^-$ final state suffers from large experimental uncertainty \cite{STAR:2009giy} so it cannot contribute to the comparison.  

For the Ref. \cite{Klusek-Gawenda:2020gwa} coupling, the data matches the cross section with about a 15 \% branching ratio. 

For the GVMD-predicted coupling, the best-fit branching ratio is unreasonably low, requiring a very large total $\rho'$ cross section.  The GVMD prediction in Eq. \ref{eq:GVMD} must be too high.  A GVMD calculation with off-diagonal elements might do better.  Alternately, a Glauber-Gribov calculation with inelastic shadowing (cross-section fluctuations) might also fit the data better
\cite{Frankfurt:2015cwa}. However, neither of these approaches is likely to lead to big enough changes to remove the tension.

Figure \ref{fig:2Dscan} shows the range of allowed values for $f_V^2/4\pi$ vs. branching ratio.  The red line shows the data, surrounded by $\pm 1$ and $\pm 2$ $\sigma$ contours.  If we assume that the branching ratio must be greater than 10\%, then a large $f_V^2/4\pi$ is preferred.  However, the branching ratio cannot be too much more than 10\%, unless the coupling is much larger than expected.  As was previously noted, small branching ratios correspond to large meson-nucleon scattering cross-sections, with even a 10\% branching ratio requiring a surprisingly large cross section.  This leaves a fairly small preferred range, with branching ratios around 15\% and coupling around the value from Ref. \cite{Klusek-Gawenda:2020gwa}.  

\begin{figure}
    \centering
    \includegraphics[width=0.48\textwidth]{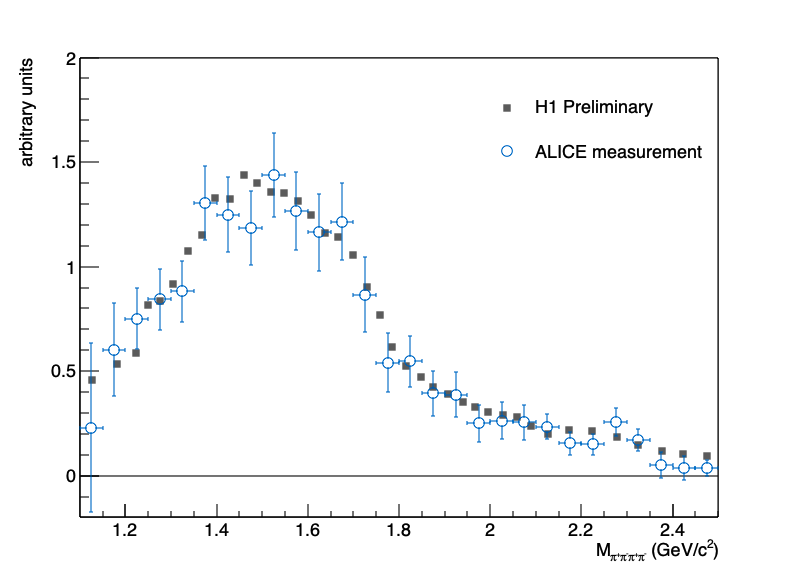}
\caption{Comparison of the invariant mass distribution of four pions from H1 collaboration (shown as gray squares)~\cite{Schmitt,H1prelim} and the ALICE collaboration~\cite{ALICE:2024kjy} in arbitrary units, normalized for easy shape comparison. Error bars are not shown for H1, but they are smaller than the error bars for ALICE.}
    \label{fig:invMass}
\end{figure}

\begin{figure}
    \centering
\includegraphics[width=0.48\textwidth]{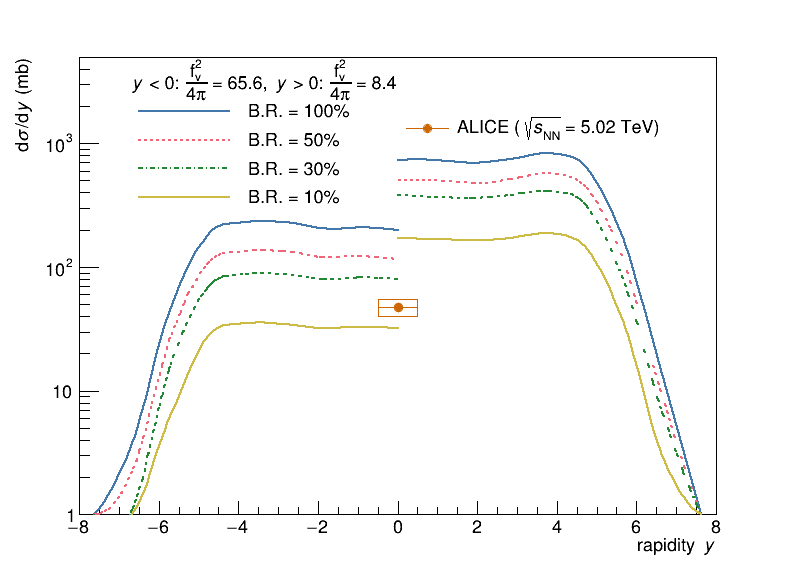}
 \includegraphics [width=0.48\textwidth]{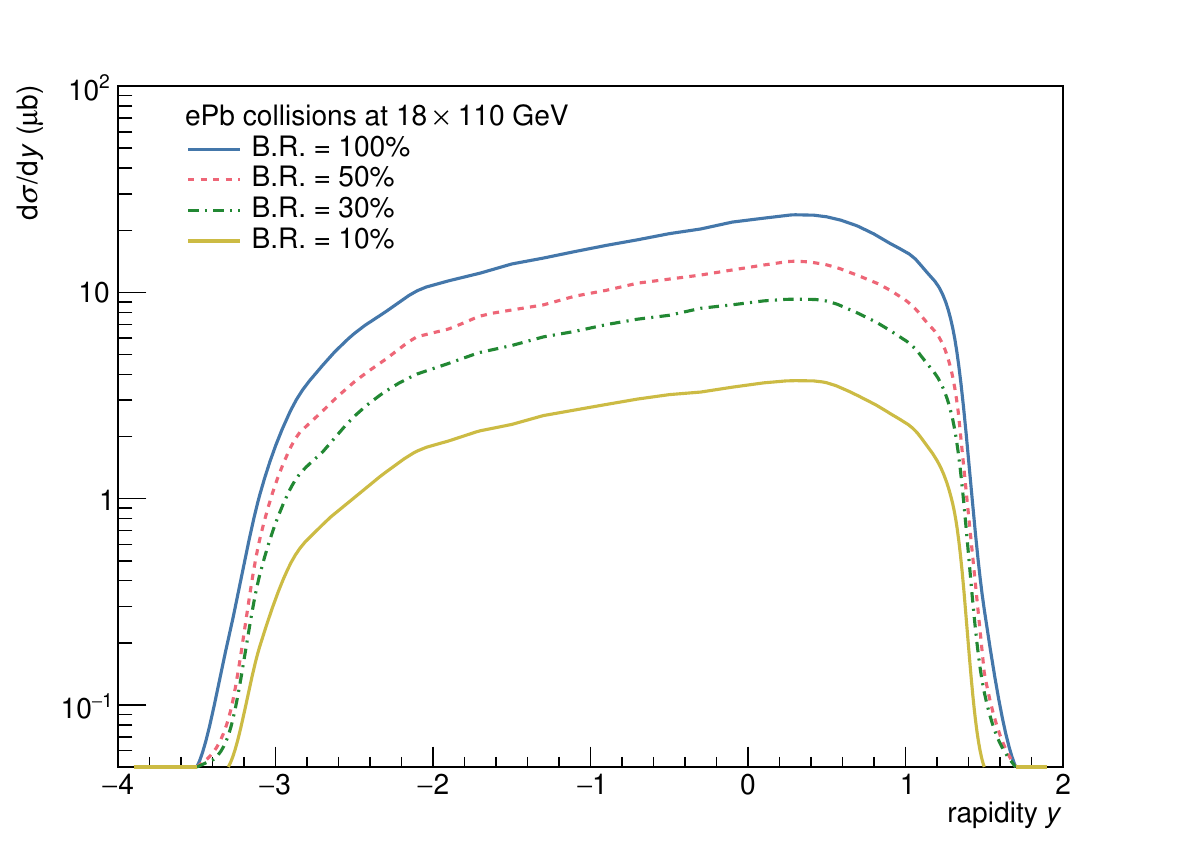}
\caption{Differential cross-section \(d\sigma/dy\) of \( \rho' \rightarrow \pi^+ \pi^- \pi^+ \pi^- \) for Pb-Pb UPCs at \( \sqrt{s_{NN}} = 5.36 \, \text{TeV} \) (top) and $e$Pb collisions at \( 18 \times 110 \, \text{GeV} \) (bottom). The curves show  different branching ratios (B.R.) of \( 100\% \), \( 50\% \), \( 30\% \), and \( 10\% \). The orange marker indicates the ALICE data for the Pb-Pb collisions at \( \sqrt{s_{NN}} = 5.02 \, \text{TeV} \).  For the UPCs(top plot), the left hand curves ($y<0$) use the coupling in Ref. \cite{Klusek-Gawenda:2020gwa}, while the right hand curves ($y>0$) use the GVMD predictions.
For ePb collisions (bottom plot), we use the coupling in Ref. \cite{Klusek-Gawenda:2020gwa}.}
    \label{fig:BRmodels}
\end{figure}

\begin{figure}
    \centering
\includegraphics[width=0.48\textwidth]{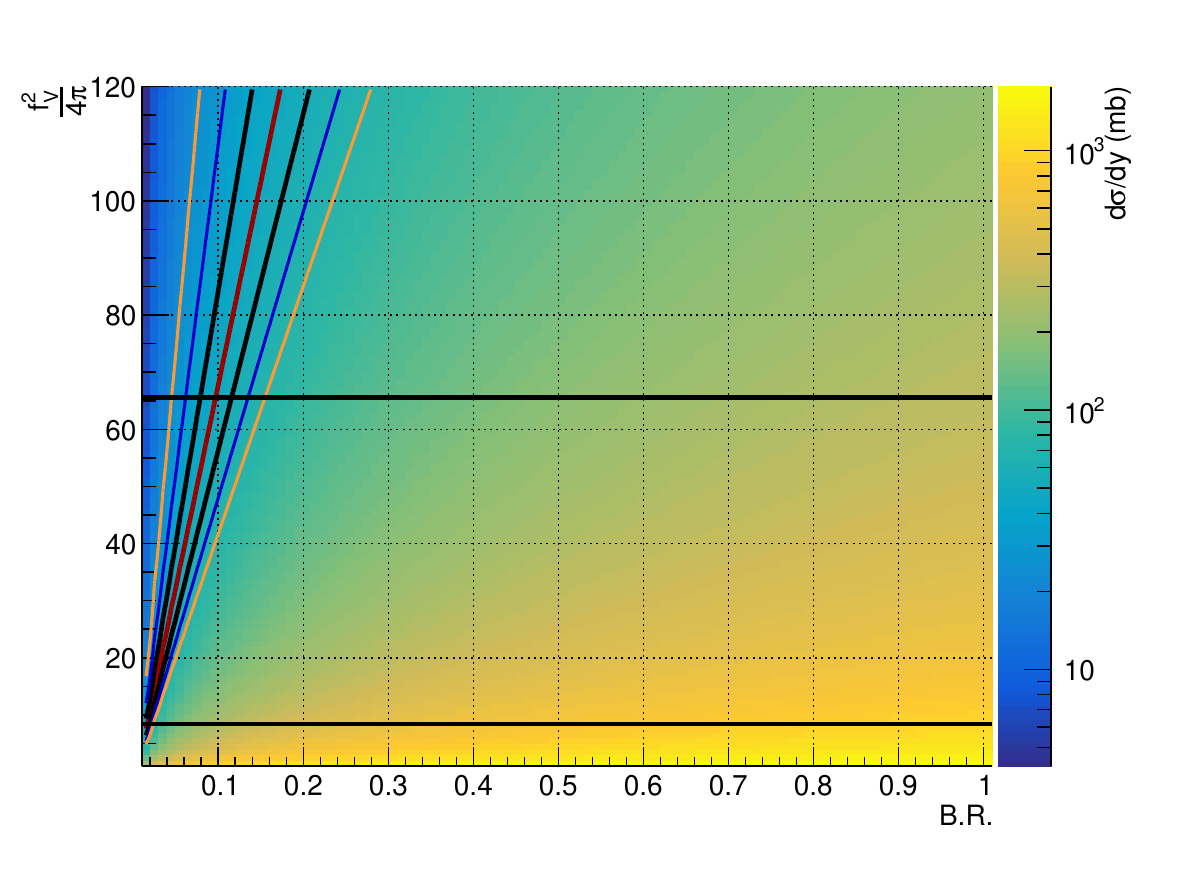}
\caption{Two-dimensional scan of the differential cross section 
$d\sigma/dy$ (in mb) as a function of the branching ratio (B.R.) 
and the coupling $f_{V}^{2}/4\pi$. 
The color map shows the interpolated cross section values on a logarithmic scale. 
Overlaid lines indicate the central value of the ALICE measurement (red solid) together with the $\pm 1\sigma$ (black), 
$\pm 2\sigma$ (blue), and $\pm 3\sigma$ (orange) contours, 
where the $\sigma$ bands represent the ALICE measurement uncertainties (stat $\oplus$ syst). 
Horizontal solid black lines mark the reference couplings 
$f_{V}^{2}/4\pi = 65.6$ and $f_{V}^{2}/4\pi = 8.4$. 
}
    \label{fig:2Dscan}
\end{figure}

If the $\rho'$ is composed of two resonances, then they are are likely to have different branching ratios and $\Gamma_{V\rightarrow ee}$.  
However, as Fig. \ref{fig:invMass} shows, the mass spectra for proton and lead targets are very similar.  This indicates that the photon-meson couplings times the branching ratio to $\pi^+\pi^-\pi^+\pi^-$ are similar for the two mesons.  Otherwise, the Glauber mapping would distort the $4\pi$ mass spectra for ion targets.  More detailed comparisons of the resonance shape and substructure should clearly show if there is one resonance or two.  High-statistics data from LHC Runs 3 and 4 \cite{Citron:2018lsq} should allow for a definitive comparison.   

We modeled the $4\pi$ final state using phase space to determine the acceptance, using simple models for four LHC detectors. The model parameters are given in Table \ref{tab:LHCacceptance}, along with the resulting detector efficiencies.   The efficiency for detecting $\rho^0\pi^+\pi^-$ would be slightly different than for this phase space distribution \cite{STAR:2009giy}, but would not alter the conclusions here. 

\begin{figure}[tb]
    \centering
    \includegraphics[width=0.48\textwidth]{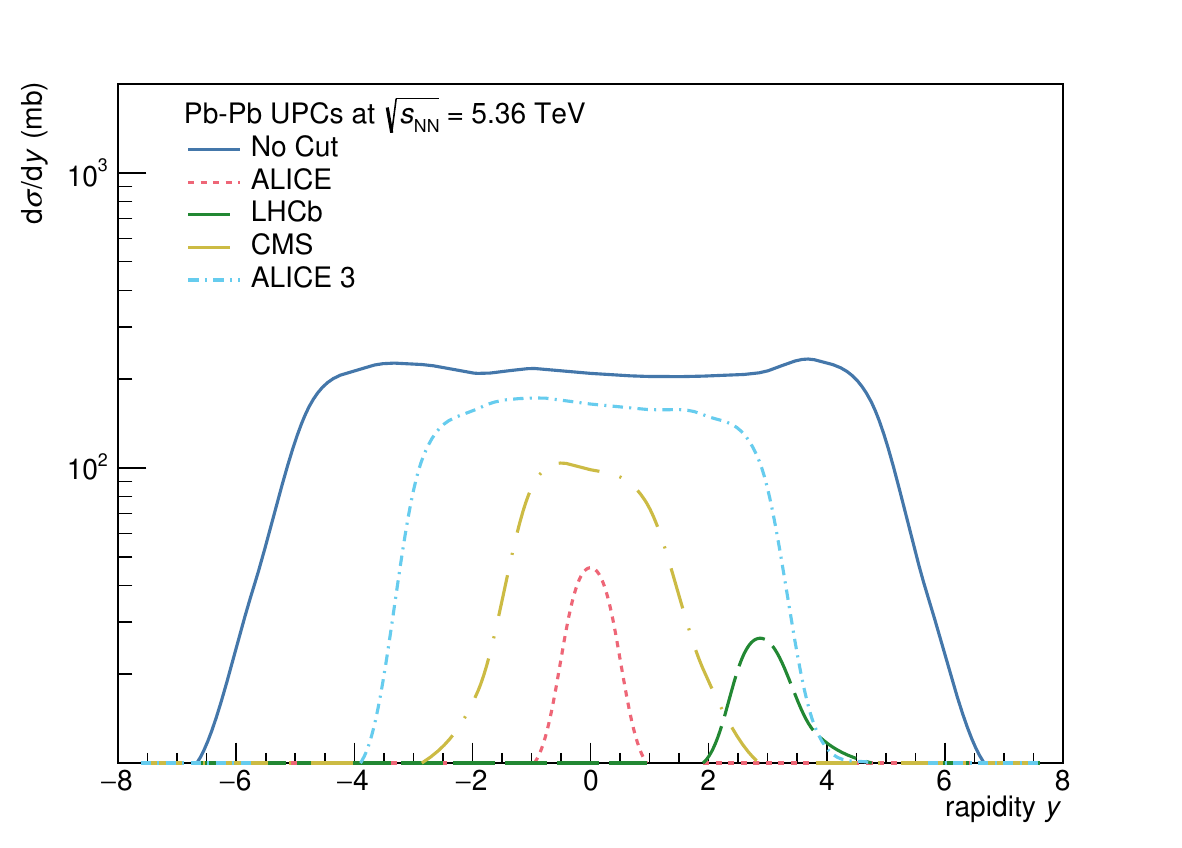}
    \caption{Differential cross-section $d\sigma/dy$ for Pb-Pb UPCs at \( \sqrt{s_{NN}} = 5.36 \, \text{TeV} \). The solid blue line represents the results with no cuts applied, while the dotted red, dashed green, and dashed-dotted yellow lines correspond to the cross sections in the ALICE~\cite{ALICE:2024kjy}, LHCb~\cite{LHCb-PAPER-2024-042}, CMS~\cite{CMS-DP-2024-011} and expected ALICE 3~\cite{arXiv:2211.02491} acceptance, respectively.}
   \label{fig:UPCacceptance}
\end{figure}

For all four detectors, the $\rho'$ signal covers a much wider rapidity range than the detectors, reducing the overall acceptance.  Figure \ref{fig:UPCacceptance} shows the signal acceptance in $d\sigma/dy$ for the different detectors. 

\begin{table}
    \centering
     \begin{tabular}{c|c|c}
     \
         Experiment & Kinematic coverage & Acceptance \\
         \hline\hline
        ALICE &$|\eta_{\pi}| < 0.9, \it{p}_{T, \pi} > \rm{0.1}\ \rm{GeV}/\it{c}$& 2.2\% \\
        \hline
        LHCb & $2.4 < \eta_{\pi} < 4.0,\ \it{p}_{T, \pi} > \rm{0.1}\ \rm{GeV}/\it{c}$& 1.8\% \\
        \hline
        CMS & $|\eta_{\pi}| < 2.4, \it{p}_{T, \pi} > {\rm 0.2}\ \rm{GeV}/\it{c}$& 12\% \\
        \hline
        ALICE 3 & $|\eta_{\pi}| < 4.0, \it{p}_{T, \pi} > {\rm 0.1}\ \rm{GeV}/\it{c}$& 43\% \\\end{tabular}
    \caption{Geometric acceptance criteria for four LHC experiments 
    \cite{ALICE:2024kjy,LHCb-PAPER-2024-042,CMS-DP-2024-011,arXiv:2211.02491}
    with their respective pseudorapidity (\(\eta_{\pi}\)) and transverse momentum (\(p_{T, \pi}\)) acceptance for charged particles, as well as the corresponding acceptances for $\rho'\rightarrow\pi^+\pi^-\pi^+\pi^-$.}
    \label{tab:LHCacceptance}
\end{table}

\section{$\rho'$ photoproduction in $ep/e\rm{A}$ collisions}

Photoproduction in $ep/e\rm{A}$ collisions adds another dimension: the photon $Q^2$.   There is no $\rho'$ data on cross sections for virtual photons, so we model the $Q^2$ evolution of the cross-section following the $\rho$ \cite{H1:2009cml,Lomnitz:2018juf}:
\begin{equation}
\sigma(W,Q^2)\!=\!\sigma(W,Q^2=0)\!\bigg(\frac{M_V^2}{M_V^2+Q^2}\bigg)^n
\end{equation}
where $n=2.09+0.73/{\rm GeV}^{-2}(M_V^2+Q^2)$.

$\rho'$ production is assumed to follow vector meson dominance, with the final state linearly polarized transverse to the beam direction at $Q^2=0$, but with an increasing longitudinal polarization as $Q^2$ rises.  The rate of this increase is not known for the $\rho'$, but we use the approach described in Ref. \cite{Lomnitz:2018juf} here, assuming that the spin matrix for the $\rho'$ is the same as for the $\rho$.    

\begin{figure*}
    \centering
    \includegraphics[width=0.48\linewidth]{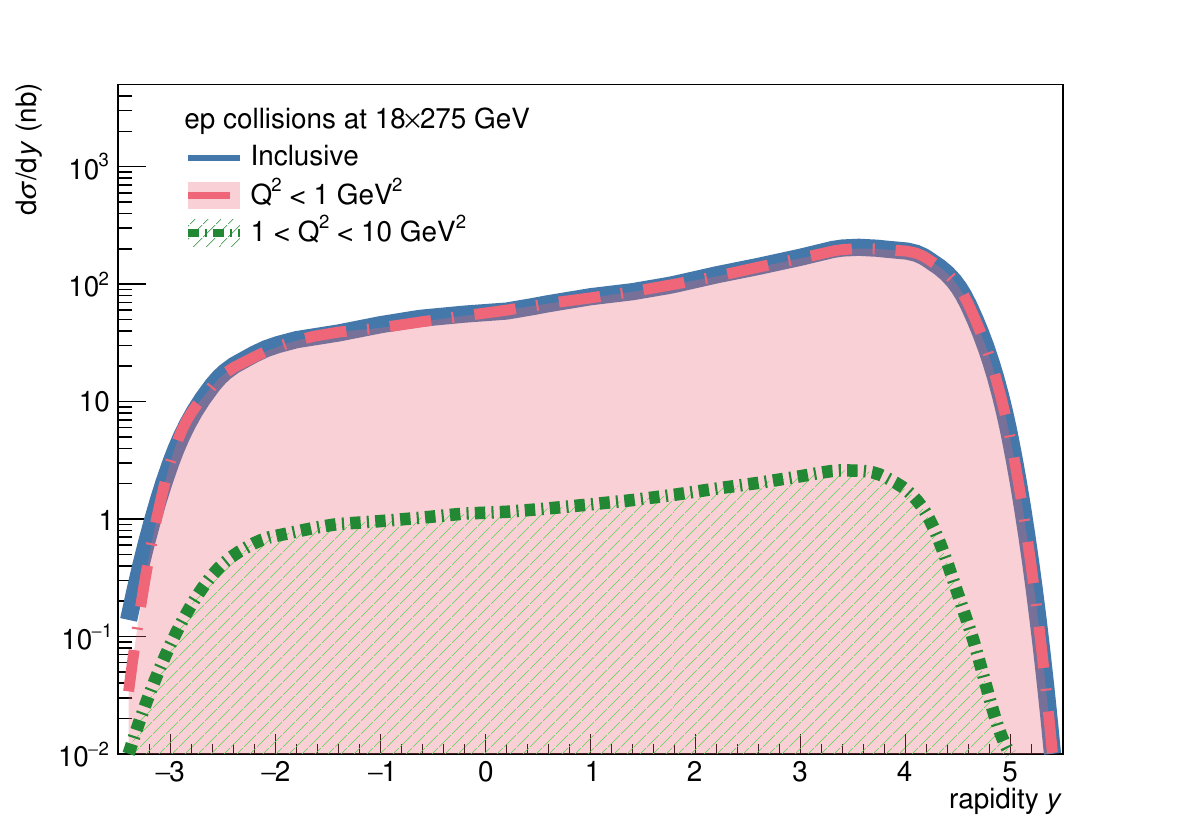}
    \includegraphics[width=0.48\linewidth]{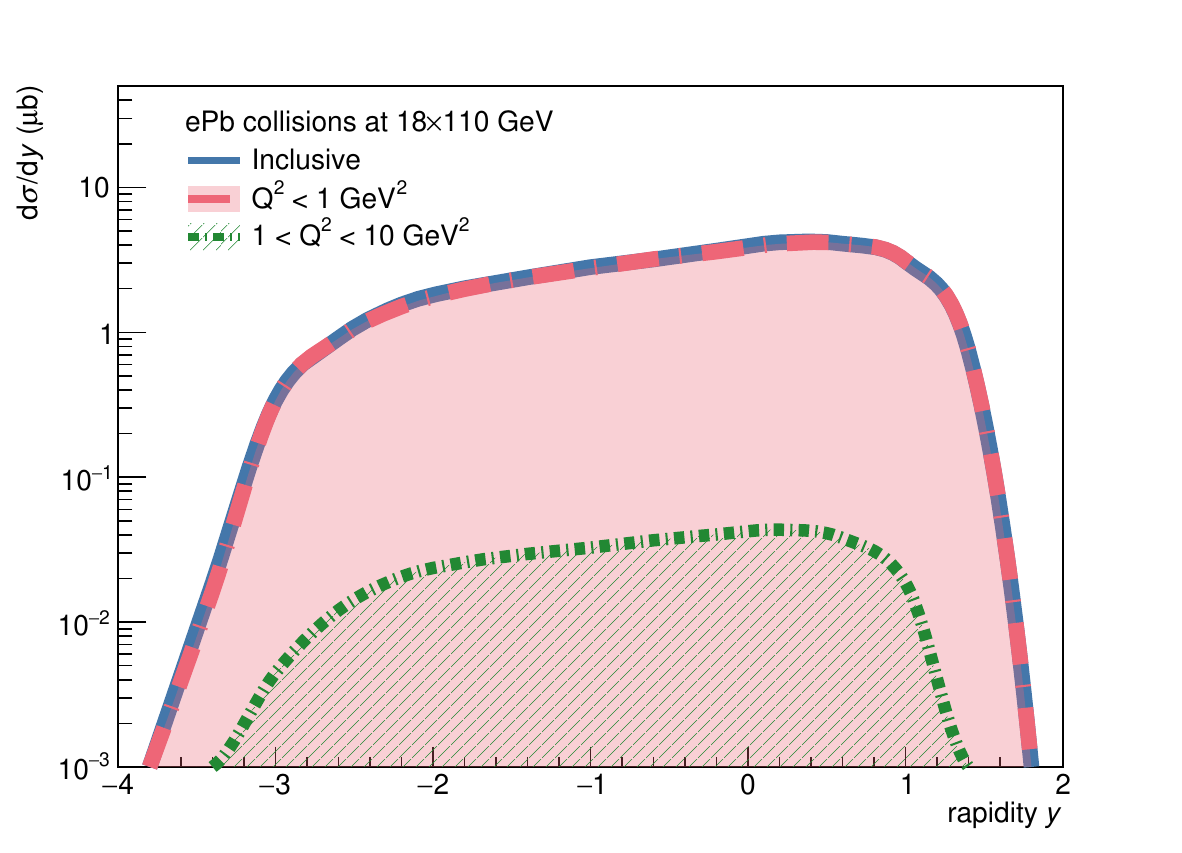}
    \caption{Cross-sections for $ep$ and $e$Pb collisions at EIC energies, presented as a function of rapidity \( y \) for $ep$ collisions at \( 18 \times 275 \, \text{GeV} \) (left) and $e$Pb collisions at \( 18 \times 110/\text{nucleon} \, \text{GeV} \) (right). The inclusive (solid line) and various \( Q^2 \) ranges (shaded areas) are indicated.}
    \label{fig:eicplot}
\end{figure*}

Simulations were performed in the eSTARlight framework \cite{Lomnitz:2018juf}.  Table \ref{tab:EICCS} shows the calculated cross sections for the top EIC energies: 18 GeV electrons colliding with  275 GeV protons or 110 GeV/nucleon lead ions.  In addition to the total ($Q^2$ integrated) cross section, cross sections are given for photoproduction  ($Q^2 < 1 \,{\rm GeV}$) and electroproduction ($1 < Q^2 < 10 \,{\rm GeV}^2$).  The total cross section is dominated by photoproduction. Figure~\ref{fig:eicplot} shows the corresponding $d\sigma/dy$.  

The cross section for $\rho'$ production in $e$Pb collisions is about 45 times that for $ep$ collisions.  This is a smaller ratio than for the $\rho$ or $\phi$, but is similar to the ratio found for $\gamma$A and $\gamma p$ collisions in Ref. \cite{Klusek-Gawenda:2020gwa}.  For $ep$ collisions, the \fourpi cross-section in $ep$ collisions is about 1/7 of the $\rho$ cross section, and about 1/3 of the $\phi$ cross section.  For $e$Pb collisions, the ratios are lower, with the $\rho'$ cross section 1/30 of the $\rho$ cross section, and half that for the $\phi$.  In both cases, the rates are high enough (7 or 1.5 billion events per 10 fb$^{-1}/A$ of integrated luminosity) to allow high-precision differential measurements.    

The rates are high enough to measure final states with small branching ratios, including $e^+e^-$.  With this measurement of $f_V$, it will be possible to determine the $4\pi$branching ratio.  This data will also make it possible to compare high and low mass \fourpi and $e^+e^-$ final states, to determine if there are one or two resonances. 

\begin{table*}
    \centering
    \renewcommand{\arraystretch}{1.2} 
    \begin{tabularx}
    {\linewidth}{c|>{\centering\arraybackslash}X>{\centering\arraybackslash}X>{\centering\arraybackslash}X>{\centering\arraybackslash}X>{\centering\arraybackslash}X}
        Collision System & $\it{Q^2}$ Range ($\rm{GeV}^{2}$) & Total Cross Section & Events for $\mathcal{L} = 10\ \text{fb}^{-1}$ ($10\ \text{fb}^{-1}/A$) & Acceptance & Acceptance with $B^0$ \\
        \hline\hline
        \multirow{3}{*}{$ep$ at $18\times275$ GeV}  
        & Inclusive & $696\ \text{nb}$ & $7.0\times 10^{9}$ & 0.39 & 0.41 \\
        & $Q^{2} < 1\ \text{GeV}^2$ & $686\ \text{nb}$ & $6.9\times 10^{9}$ & 0.39 & 0.41 \\
        & $1 < Q^{2} < 10\ \text{GeV}^2$ & $10.0\ \text{nb}$ & $1.0\times 10^{8}$ & 0.52 & 0.53 \\
        \hline
        \multirow{3}{*}{$e$Pb at $18\times110/A$ GeV}  
        & Inclusive & $31.3\ \mu\text{b}$ & $1.5\times 10^{9}$ & 0.73 & 0.73 \\
        & $Q^{2} < 1\ \text{GeV}^2$ & $30.6\ \mu\text{b}$ & $1.4\times 10^{9}$ & 0.73 & 0.73 \\
        & $1 < Q^{2} < 10\ \text{GeV}^2$ & $0.639\ \mu\text{b}$ & $3.1\times 10^{6}$ & 0.76 & 0.76 \\
        \hline
    \end{tabularx}
    \caption{Projected cross sections and event rates for $ep$ and $e$Pb collisions at EIC energies, based on the current EPIC detector design, for different $Q^{2}$ ranges. The integrated luminosities ($\mathcal{L}$) used are $10\ \text{fb}^{-1}$ for $ep$ and $10\ \text{fb}^{-1}/A$ for $e$Pb collisions. Acceptance values are based on simulations using the current detector design.}
    \label{tab:EICCS}
\end{table*}

\begin{figure*}
    \centering
    \includegraphics[width=0.48\linewidth]{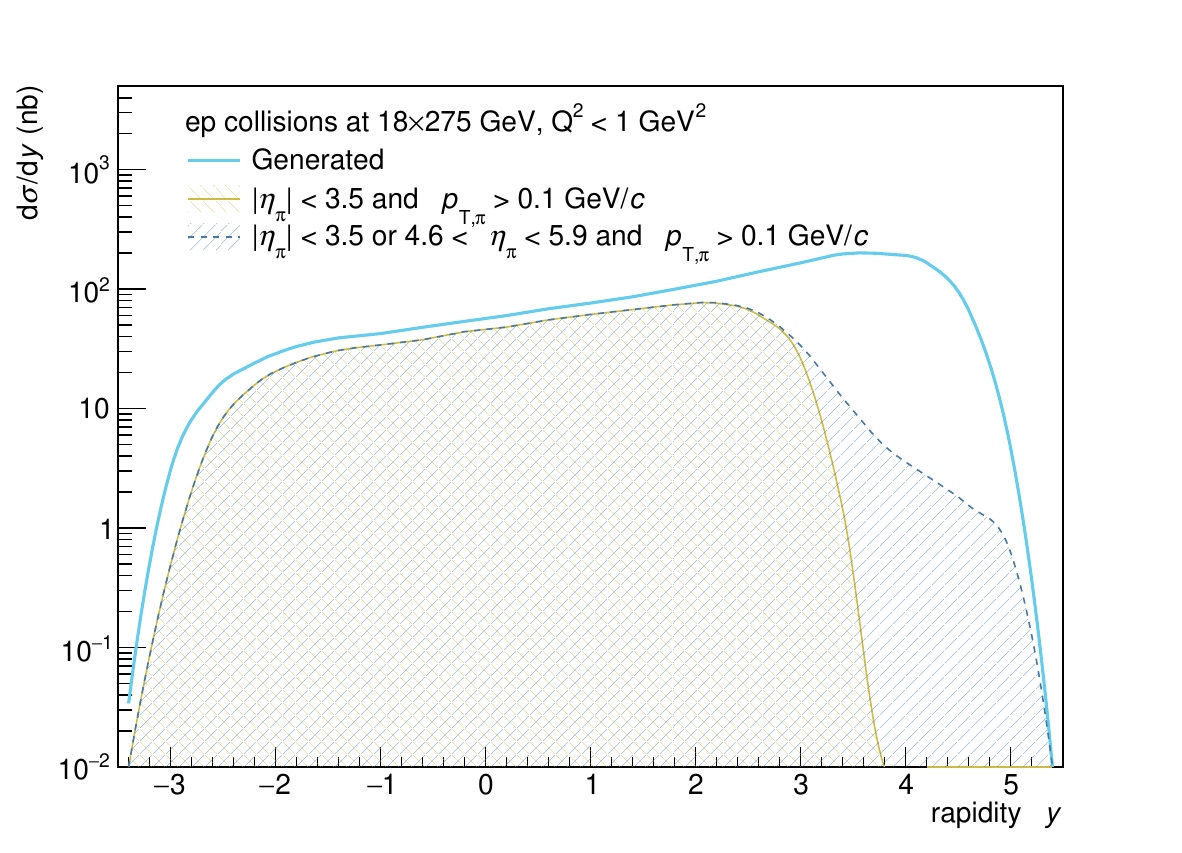}
    \includegraphics[width=0.48\linewidth]{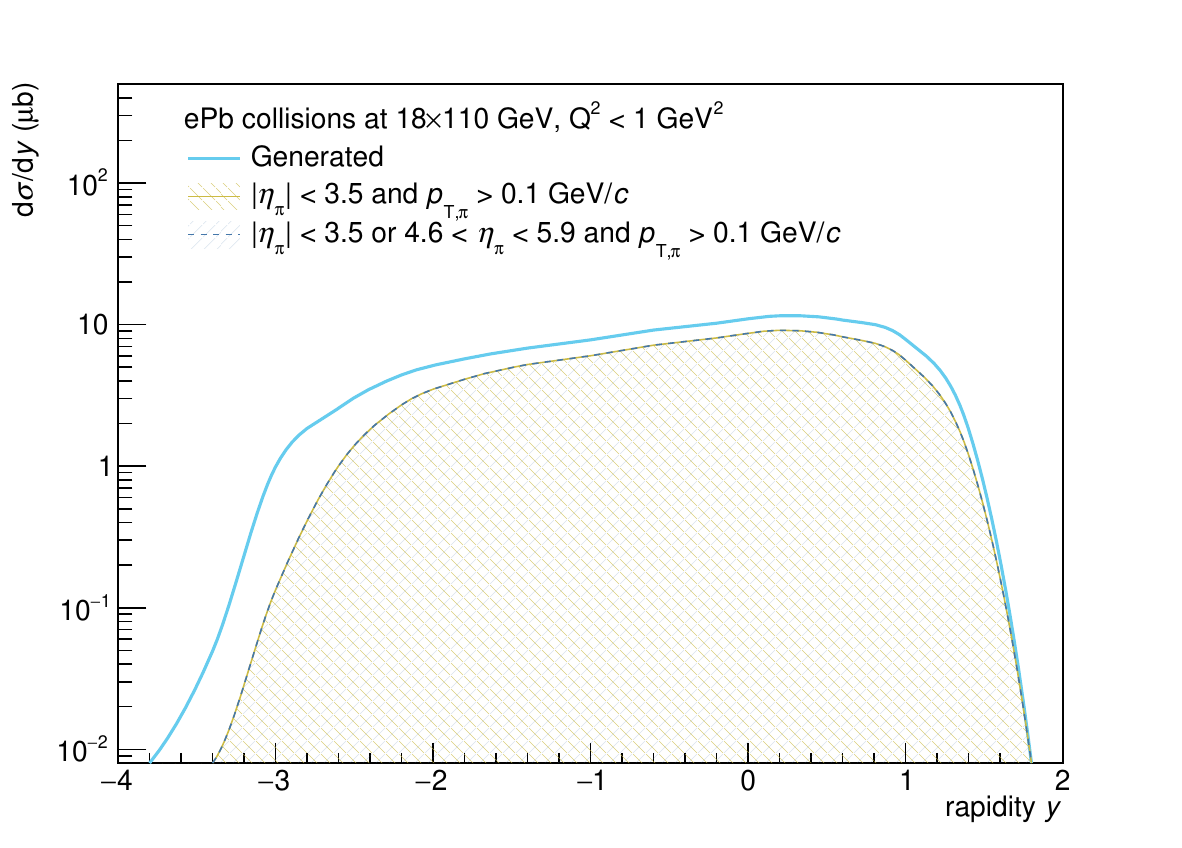}
    \caption{Accessible cross sections for $ep$ collisions at \( 18 \times 275 \, \text{GeV} \) (left) and $e$Pb collisions at \( 18 \times 110/\text{nucleon} \, \text{GeV} \) (right), with specific kinematic cuts based on the current EPIC detector tracker design. The blue shaded area with dotted lines represents the coverage including the B0 tracker.}
    \label{fig:eicplot_ePIC}
\end{figure*}

The reconstruction efficiency for the ePIC detector is shown in Fig. \ref{fig:eicplot_ePIC}.  ePIC is modeled with two tracking components: a central barrel tracker sensitive to $|\eta|<3.5$ and the B0 detector, which covers $4.6 < \eta < 5.9$.  For both components, we required track $p_T > 100$ MeV/c, but this cut had little effect on the $\rho'$ efficiency.  Fig. \ref{fig:eicplot_ePIC} show the geometric detection efficiency for photoproduction events.  We do not consider detector inefficiency within the geometric acceptance.  This inefficiency should be small.

The efficiency is high for $|\eta|<2.5$, but falls off at larger $|y|$, as expected.  The B0 detector plays an important role for $ep$ collisions in the region $y >3$; otherwise the efficiency would be near zero for $y >3.5$.  Even though the efficiency is fairly low in this region, the rates are high enough that high statistics data should be achievable.  

The $\rho'$ rapidity is related to the target Bjorken$-x$
\begin{equation}
x= \frac{M_{\rho'}}{2m_p\gamma} \exp{(-y)},
\end{equation} 
where $m_p$ is the proton mass and $\gamma=292$ is the Lorentz boost of the proton beam at maximum EIC energy.  For UPCs, a similar relation holds, except that there is a $\pm$ sign in the exponential, since the photon direction is unknown.

The central detector cutoff, $\eta=3.5$, corresponds roughly to $y\approx 3.5$ (as can be seen in Fig. \ref{fig:eicplot_ePIC}), or $x\approx 10^{-4}$.  For $ep$ collisions, covering the full range of Bjorken$-x$ requires acceptance out to rapidity $\approx 5$, corresponding to $x\approx2 \times 10^{-5}$.  As Fig. \ref{fig:eicplot_ePIC} shows, the B0 tracker is needed to cover this low$-x$ region.  Nuclear shadowing (beyond that present in the Glauber calculation) or saturation would manifest itself as a reduction in cross-section with decreasing $x$, {\it i. e. with increasing rapidity} at low/moderate $Q^2$ \cite{Mantysaari:2017slo}.  There would also be changes in $d\sigma/dt$ 
\cite{Accardi:2012qut}. The rates are high enough to permit these measurements,  even at the lowest $x$.  

In short, the  $\rho'$ is copiously produced and relatively easy to reconstruct, showing promise for use in saturation studies.  The rates given in Table \ref{tab:EICCS} assume a branching ratio of 100\%.  For a more likely branching ratio around 10\%, the rates are reduced by a factor of about 8, but are still high enough for high-statistics measurements.  

As with UPCs, the main uncertainties are from the unknown $f_V$ and branching ratios.  Fortunately, the coupling to $e^+e^-$ can be directly determined experimentally, by measuring that final state.  Other uncertainties are much smaller, making the $\rho'$ an excellent medium-mass candidate for mapping out shadowing as a function of $Q^2$.

\section{Discussion and conclusions}

We have calculated the cross section for $\rho'\rightarrow\pi^+\pi^-\pi^+\pi^-$ on ion targets, using proton-target data for photoproduction on proton targets as input.   The resulting ion-target cross sections depend on the photon-meson couplings and the branching ratio for the $\rho'$ to decay to $4\pi$.   The couplings predicted using a GVMD model lead ion-target cross-sections that are high enough to require implausibly small branching ratios.  Using the coupling from Ref. \cite{Klusek-Gawenda:2020gwa},  ALICE data on $\rho'\rightarrow\pi^+\pi^-\pi^+\pi^-$, prefers branching ratios around 15\%.   This estimate uses a Glauber calculation, but small deviations will not have a significant effect on this conclusion.  There is some tension between the cross-section ratio and the inferred meson-nucleon scattering ross section.

With LHC Run 3/4 and EIC data, it should also be possible to measure the coupling to $e^+e^-$, using higher $p_T$ $e^+e^-$ pairs from incoherent photoproduction, to avoid backgrounds from $\gamma\gamma\rightarrow e^+e^-$.   It should also be possible to investigate whether a GVMD model with off-diagonal intermediate states might better predict the photon-meson couplings \cite{Kobayashi:1973ep,Fraas:1977ef,Frankfurt:2003wv}.

We also made predictions for $\rho'\rightarrow\pi^+\pi^-\pi^+\pi^-$ at the proposed U. S. EIC. The production rates are high, and the meson is an easy-to-reconstruct probe of nuclear structure.  The $\rho'$ is intermediate in mass between the $\rho^0$ and the $J/\psi$, so it should show significant saturation, but also be amenable to pQCD calculations.   These features should also make it an attractive target at the proposed Chinese electron-ion collider, EiCC \cite{Chen:2020ijn} or the proposed LHeC collider \cite{LHeC:2020van}.  For the LHeC, excellent forward instrumentation is critical to be able to observe photoproduction and electroproduction at the highest energies, corresponding to the lowest Bjorken$-x$.  The wide $W_{\gamma p}$ range at the LHeC data will better constrain the energy dependence of $4\pi$ photoproduction. 

The approach developed here, comparing $\gamma p$ and $\gamma$A collisions may be applicable for determining absolute branching ratios for other mesons,  as long as the photon-meson coupling is known. 

This work is supported in part by the U.S. Department of Energy, Office of Science, Office of Nuclear Physics, under contract numbers DE-AC02-05CH11231 and DE-FG02-96ER40991.   

\bibliographystyle{apsrev4-1} 
\bibliography{inco}

\end{document}